# An Ultra-Low Latency, End-to-End Streaming Speech Synthesis Architecture via Block-Wise Generation and Depth-Wise Codec Decoding

**Author:**

Tianhui Su[1], Tien-Ping Tan[1,*], Salima Mdhaffar[2], Yannick Estève[2], Aghilas Sini[3]

*Corresponding author: tienping@usm.my

[1] School of Computer Sciences, Universiti Sains Malaysia, 11800 USM, Penang, Malaysia

[2] 74 Rue Louis Pasteur, LIA, Avignon University ,84029 Avignon, France

[3] Av. Olivier Messiaen, LIUM, Le Mans University ,72085 Le Mans, France

# Abstract

Real-time speech synthesis requires a rigorous balance between inference latency and acoustic fidelity to support interactive human-computer applications. Conventional continuous text-to-speech pipelines necessitate computationally intensive neural vocoders to reconstruct phase information, creating a significant streaming bottleneck. Furthermore, regression-based speech synthesis acoustic modeling frequently induces spectral over-smoothing artifacts that degrade perceptual quality. To address these limitations, this paper proposes a novel end-to-end non-autoregressive architecture optimized for ultra-low latency block-wise generation, which directly models the highly compressed discrete latent space of the Mimi neural audio codec. By integrating a modified FastSpeech 2 backbone with a progressive depth-wise sequential decoding strategy, the proposed architecture dynamically conditions 32 layers of residual vector quantization codes. This mechanism effectively resolves phonetic alignment degradation and manages the structural complexity of high-fidelity discrete representations without introducing temporal autoregressive overhead. Experimental evaluations conducted on English and Malay single-speaker datasets validate the language-independent deployment capability of the proposed architecture. Compared to conventional continuous regression models, the proposed architecture demonstrates quantitative improvements in fundamental voicing accuracy and effectively mitigates high-frequency spectral degradation. Furthermore, it achieves ultra-low latency inference, translating to a 10.6-fold absolute acceleration over conventional cascaded continuous pipelines. Crucially, the architecture achieves an average time-to-first-byte latency of 48.99 milliseconds, falling significantly below the human perception threshold for real-time interactive streaming. The results firmly establish the proposed discrete architecture as a highly optimized solution for deploying high-quality, real-time streaming speech interfaces in resource-constrained environments.

## 1. Introduction

Speech synthesis, or text-to-speech (TTS) conversion, is the artificial production of human speech from text or phonological representations. The increasing demand for real-time interactive applications has prioritized the development of streaming synthesis systems capable of immediate responsiveness (Jampala et al., 2024; Kaur & Singh, 2023). Unlike conventional offline generation, streaming architectures require the incremental synthesis of speech segments with minimal algorithmic latency to maintain the natural turn-taking rhythm of conversational dynamics (Juvela et al., 2019).

Traditional neural synthesis systems typically employ a cascaded two-stage paradigm, wherein an acoustic model generates intermediate continuous representations from text before a dense neural vocoder reconstructs the time-domain waveform (Kong et al., 2020; J. Shen et al., 2018). Although non-autoregressive acoustic models (Ren, Hu, et al., 2022) have substantially accelerated initial feature generation, the inherent reliance on continuous Mel-spectrograms presents fundamental structural limitations. These continuous representations discard critical phase information, unconditionally necessitating computationally heavy phase-estimation networks to synthesize the final audio (Ueno & Kawahara, 2022). This structural dependency introduces severe algorithmic latency and processing overhead, creating an execution bottleneck for efficient streaming implementations (Tan et al., 2024).

Furthermore, formulating acoustic modeling as a continuous regression task optimized via mean squared error frequently results in spectral over-smoothing (Ren, Tan, et al., 2022). This phenomenon forces the model to predict the statistical mean of the target distribution, yielding degraded speech that fundamentally lacks high-frequency spectral textures and transient phonetic details.

To circumvent the computational burden of heavy neural vocoders and the acoustic limitations of continuous regression, recent research has pivoted toward discrete neural audio codecs (Défossez et al., 2024). These advanced codecs compress raw audio into highly compact discrete latent tokens utilizing multi-layered residual vector quantization (Zeghidour et al., 2022). However, seamlessly integrating these discrete representations into a streaming non-autoregressive architecture presents significant architectural challenges. The hierarchical nature of residual quantization, which often involves extensive codebook depths, makes simultaneous parallel prediction highly unstable and prone to phonetic alignment collapse. Conversely, existing discrete language modeling approaches rely on strict temporal autoregressive generation. By flattening the hierarchical layers into incredibly long one-dimensional sequences, these autoregressive models suffer from prohibitive inference latency, rendering them fundamentally unsuitable for instantaneous streaming applications (Dang et al., 2024).

In this paper, we propose an efficient end-to-end language independent single speaker streaming speech synthesis architecture. By directly mapping linguistic features into the discrete hierarchical latent space of the pre-trained neural codec via a novel depth-wise sequential decoding strategy, the model completely eradicates the heavy continuous phase-estimation pipeline while preserving fine-grained acoustic fidelity.

The primary contributions of this work are summarized as follows:

End-to-End Streaming TTS Architecture: We propose an end-to-end architecture that seamlessly integrates a non-autoregressive linguistic encoder with a discrete neural audio decoder. This structural paradigm bypasses the traditional cascaded bottleneck, enabling direct and exceptionally fast text-to-waveform generation suitable for ultra-low latency streaming environments.

Depth-Wise Sequential Decoding Strategy: To resolve the phonetic alignment degradation inherent in naive parallel prediction, we introduce a progressive feature fusion mechanism. This strategy recursively conditions the higher-order residual quantizers upon the semantic foundations established by the initial layers strictly within the feature depth of individual frames, effectively bridging the representational gap between semantic accuracy and acoustic texture without introducing temporal delays.

Language Independent Structural Robustness: We validate the proposed discrete modeling paradigm across distinct linguistic structures utilizing English and Malay benchmarks. The empirical results demonstrate that the architecture achieves fundamental voicing accuracy, viable speech intelligibility, and a significant real-time generation speedup across diverse phonological systems without requiring dataset-specific modifications.

## 2. Related Work

In this section, we discuss the evolutionary trajectory of acoustic modeling and the critical architectural shifts that have shaped modern streaming speech synthesis. We first review the progression from autoregressive to non-autoregressive (NAR) architectures, highlighting the persistent computational bottlenecks associated with conventional continuous Mel-spectrograms and neural vocoders. Subsequently, we examine contemporary streaming TTS architectures, analyzing their inherent latency-fidelity trade-offs and structural constraints. Finally, we explore the recent paradigm shift towards discrete neural audio codecs and Residual Vector Quantization (RVQ), identifying the structural challenges of modeling hierarchical discrete representations in parallel, which ultimately motivates the design of our proposed end-to-end architecture.

## 2.1 Non-Autoregressive TTS and Acoustic Modeling Limitations

Deep learning has fundamentally transformed TTS synthesis, evolving from sequential autoregressive (AR) architectures to highly efficient non-autoregressive (NAR) architecture. Early AR models, such as Tacotron 2 (J. Shen et al., 2018) and Transformer TTS (N. Li et al., 2019), established a high standard for prosodic naturalness by utilizing sequence-to-sequence attention mechanisms. However, these models suffered from inherent inference latency due to their frame-by-frame generation process, which exhibits an $O(N)$ computational complexity scaling linearly with the output sequence length. To address this bottleneck, NAR models like FastSpeech (Ren et al., 2019), FastSpeech 2 (Ren, Hu, et al., 2022), and LightSpeech (Luo et al., 2021) introduced parallel generation mechanisms. By leveraging duration predictors and variance adaptors, these architectures successfully decoupled acoustic generation from temporal autoregression, achieving massive inference speedups and significantly reducing algorithmic delay. To contextualize this evolution, Table 1 summarizes the architectural properties, inference paradigms, and speed profiles of these and other prominent non-autoregressive acoustic models developed in recent years.

| **Model** | **Acoustic Target** | **Vocoder Dependency** | **Inference Speed Profile** | **paradigm** | **Key Real-Time Bottleneck** |
|---|---|---|---|---|---|
| **FastSpeech** | Continuous Mel-spectrogram | High (eg: HiFi-GAN) | High Speed | Parallel, duration-based | Phase-reconstruction computational load |
| **FastSpeech 2** | Continuous Mel-spectrogram | High (eg: HiFi-GAN) | High Speed | Parallel, prosody-aware | Vocoder computational cost |
| **VITS** | Latent / waveform | None (Integrated) | High Speed | End-to-end flow + latent variables | Complex adversarial optimization overhead |
| **Glow-TTS** | Continuous mel-spectrogram | High | High Speed | Flow modeling | alignment complexity |
| **Grad-TTS** | Continuous mel-spectrogram | High | High Speed | Diffusion-based generative modeling | Iterative denoising sampling latency |
| **NaturalSpeech 2** | Continuous Latent Vectors | High (Diffusion) | Variable | Latent diffusion model | Iterative diffusion sampling can |

| | | | | | lead to execution delays |
|---|---|---|---|---|---|
| **LightSpeech** | Continuous Mel-spectrogram | High (eg: HiFi-GAN) | High Speed | Lightweight, distillation-based Non-AR | Vocoder + reduced expressiveness due to compact model |
| **E2 TTS** | Continuous (implicit latent → waveform) | None (End-to-End) | Variable | Flow matching / ODE-based | ODE solver / numerical integration cost |

Table 1 Performance Summary of Prominent Non-Autoregressive TTS Architectures

Despite these architectural advancements in the acoustic modeling stage, the majority of existing NAR systems remain tethered to a two-stage cascaded paradigm: predicting continuous Mel-spectrograms followed by a distinct neural vocoder (e.g., HiFi-GAN (Kong et al., 2020) or WaveGlow (Prenger et al., 2019)) for phase estimation and waveform reconstruction. From a systems engineering perspective, this traditional approach presents critical limitations for practical deployment. Specifically, the absolute dependency on heavy neural vocoders introduces substantial computational overhead. State-of-the-art vocoders rely on dense transposed convolutions that consume massive floating-point operations and memory bandwidth, thereby creating a severe execution bottleneck for streaming applications on resource-constrained edge devices (Tan et al., 2024; Ueno & Kawahara, 2022). Furthermore, while architectures like Glow-TTS (Kim et al., 2020) and VITS (Kim et al., 2021) integrate normalizing flows to bypass external vocoders, their complex adversarial and flow-based optimization mechanisms still impose notable computational overhead. Similarly, emerging end-to-end continuous models like E2 TTS (Eskimez et al., 2024) frequently rely on flow-matching algorithms where inference speed remains tightly bound to the numerical integration cost of ordinary differential equation solvers, complicating strict ultra-low latency deployments.

Additionally, the acoustic modeling of continuous features via standard regression objectives fundamentally suffers from spectral over-smoothing artifacts. Because speech is a highly multi-modal distribution, regression-based optimization forces the network to predict the statistical mean of potential acoustic variations, leading to the loss of high-frequency textures and resulting in a muffled or buzzy output quality (Nippert, 2025; Ren, Tan, et al., 2022). While recent generative paradigms have successfully mitigated this over-smoothing by modeling the exact data distribution, as exemplified by generative adversarial networks (Saito et al., 2019) and denoising diffusion probabilistic models (C. Wang et al., 2023) such as Grad-TTS (Popov et al., 2021) and NaturalSpeech 2 (K. Shen et al., 2024), their iterative sampling procedures incur prohibitive computational costs. These extensive execution delays fundamentally violate the strict algorithmic latency requirements demanded by real-time streaming speech synthesis.

**2.2 Streaming TTS Architectures**

Real-time conversational agents require streaming architectures capable of incremental synthesis, prioritizing a minimal "time-to-first-byte" (TTFB) to ensure seamless human-machine interaction (Borsos et al., 2023). Standard streaming architectures typically employ chunk-based or block-wise processing mechanisms paired with restricted look-ahead windows (Ellinas et al., 2021; Ma et al., 2020). By processing text input in discrete linguistic chunks, these systems aim to generate corresponding acoustic features synchronously. To contextualize these specific processing strategies and their inherent structural bottlenecks, Table 2 outlines highly influential streaming speech synthesis architectures developed for interactive applications.

| Architecture /Method | Streaming Strategy | Algorithmic Latency Profile | Primary Structural Bottleneck |
| --- | --- | --- | --- |
| **Streaming Tacotron** | Chunk-based autoregressive attention | High (Chunk-level context) | Sequential acoustic frame generation incurs temporal delay |
| **SyncSpeech** | Dual-stream synchronized buffering | High (Sentence-level context) | High context-dependency can increase buffering delay |
| **StreamVITS** | Causal flow-based generation | Medium (Chunk-dependent) | Overlap-add chunk buffering is often associated with latency |
| **LiveSpeech** | Autoregressive codec token prediction | Medium (Token-level) | Temporal autoregressive generation may introduce overhead |
| **StreamVoice** | Block-wise zero-shot synthesis | Low (Frame-level) | Often reliant on continuous intermediate features |

Table 2 Performance Summary of Streaming Speech Synthesis Architectures

  Achieving high acoustic fidelity under strict streaming constraints involves a delicate structural trade-off between algorithmic context dependency and computational processing speed. Reducing the look-ahead context directly degrades prosodic coherence because the model fundamentally lacks knowledge of the future sentence structure (Chen et al., 2021). To combat this limitation within autoregressive architectures, early chunk-based methods like Streaming Tacotron (Ellinas et al., 2021) enforced monotonic alignments to favor continuous output, albeit still suffering from temporal generation delays (Zhang et al., 2024). Subsequent structural innovations sought to bypass autoregression entirely. For instance, SyncSpeech (Sheng et al., 2025) utilized dual-stream transformer mechanisms to synchronize acoustic feature generation without excessive buffering. As generative paradigms advanced, architectures such as StreamVITS (Bai et al., 2025; Kim et al., 2021) implemented causal flow-based generation to adapt continuous latent variables explicitly for real-time output.

  Nevertheless, the cascaded vocoder remains a major impediment to pure streaming efficiency. Continuous feature vocoders (like CNN-based HiFi-GAN) require overlapping acoustic frames (receptive fields) to ensure phase continuity at the chunk boundaries (Juvela et al., 2019). This overlap-add requirement forces the system to buffer additional future frames before generating the current audio chunk, inherently increasing the algorithmic delay. More recently, discrete acoustic modeling has emerged to circumvent traditional continuous bottlenecks. Architectures such as LiveSpeech (Dang et al., 2024) employ autoregressive codec token prediction to achieve token-level streaming, whereas StreamVoice (Z. Wang et al., 2024) utilizes block-wise zero-shot synthesis. However, temporal autoregressive token generation inevitably introduces processing overhead, and block-wise mechanisms frequently remain reliant on intermediate continuous alignment features. Consequently, there is a compelling structural need for unified, end-to-end architectures that can map linguistic streams directly into high-fidelity waveforms, completely bypassing the intermediate continuous representations and their associated latency penalties.

**2.3 Discrete Neural Audio Codecs and Vector Quantization**

To circumvent the inherent limitations of continuous Mel-spectrograms and dense neural vocoding, the speech processing community has recently pivoted towards discrete latent representations derived from neural audio codecs. Building upon the foundational principles of Vector Quantized-Variational Autoencoders (VQ-VAE) (van den Oord DeepMind et al., 2017), modern neural codecs such as SoundStream (Zeghidour et al., 2022), EnCodec (Défossez et al., 2022), and Mimi (Défossez et al., 2024) compress raw audio waveforms into highly compact, discrete tokens. Furthermore, advancements in technologies such as Descript Audio Codec (Kumar et al., 2023) have significantly improved the mitigation of quantization artifacts. These codecs utilize Residual Vector Quantization (RVQ), which cascades multiple quantizer layers to encode audio hierarchically. The initial layers capture primary semantic and prosodic structures, while deeper layers progressively encode fine-grained acoustic textures and high-frequency details. This discrete representation not only preserves phase information better than continuous spectrograms but also operates at a significantly reduced frame rate (e.g., 12.5 Hz to 50 Hz), offering massive advantages for computational efficiency (Langman et al., 2025). To contextualize this representational shift, Table 3 summarizes the target bitrate ranges and temporal efficiencies of prominent high-fidelity neural audio codecs.

| Codec | Quantization | Target Bitrate | Temporal Resolution | Key Advantage for Synthesis System |
|---|---|---|---|---|
| **SoundStream** | Residual Vector Quantization | 3.0-18.0 kbps | 50 Hertz | Early unified end-to-end neural audio codec |
| **EnCodec** | Residual Vector Quantization | 1.5-24.0kbps | ~50-75 Hertz | High-fidelity compression across variable bitrates |
| **Descript Audio Codec** | Improved Residual Quantization | 8.0 kbps | 86 Hertz | Reduction of quantization artifacts |
| **Mimi Codec** | Semantic and Acoustic Dual-RVQ | 1.1 kbps | 12.5 Hertz | Ultra-low frame rate suitable for streaming |

Table 3 Performance Summary of High-Fidelity Neural Audio Codecs

This representational shift has catalyzed the emergence of a new class of generative speech models that formulate synthesis purely as a discrete language modeling task. Prominent architectures such as VALL-E (C. Wang et al., 2023), SPEAR-TTS (Kharitonov et al., 2023), and AudioLM (Borsos et al., 2023) autoregressively predict codec tokens utilizing large transformer architectures. These models demonstrate unprecedented zero-shot synthesis capabilities and excellent mitigation of acoustic over-smoothing. However, because these autoregressive architectures rely on strict temporal sequential generation, frequently flattening the extended time axis and the multi-layered residual quantization codes into a singular prolonged sequence, they may introduce significant inference latency. This computational overhead can render them less suitable for practical deployment in strict real-time streaming scenarios (Dang et al., 2024).

Conversely, adapting these discrete representations to parallel non-autoregressive architectures presents a profound structural challenge. The hierarchical complexity of residual vector quantization codes inherently involves multiple parallel layers per temporal frame, rendering simultaneous modeling exceedingly difficult. Existing parallel discrete architectures, such as SoundStorm (Borsos et al., 2023) or MaskGCT (Y. Wang et al., 2024), rely on iterative masked token modeling. While these architectures achieve parallel generation, their iterative refinement steps still consume significant execution time. Currently, non-iterative and fully parallel streaming architectures capable of modeling deep multi-layered discrete targets without compromising inference speed remain scarce in the literature (Han et al., 2025).

This gap highlights a strict necessity for efficient decoding mechanisms capable of bridging non-autoregressive temporal modeling with the hierarchical conditioning of discrete codecs. To resolve this structural conflict, this paper proposes a depth-wise sequential decoding strategy integrated within a streaming non-autoregressive architecture. While maintaining absolute temporal parallelism across acoustic frames, the discrete tokens within a single frame are predicted sequentially across the quantization depth. This progressive feature fusion explicitly conditions each subsequent quantizer layer on the accumulated acoustic memory of the preceding layers. Consequently, this hierarchical cascade aims to restore fine-grained acoustic textures while securely preserving the ultra-low real-time factor demanded by streaming deployments.

**2.4 Integrating Non-Autoregressive TTS with Neural Audio Codecs**

The integration of non-autoregressive acoustic models with discrete neural audio codecs represents the current frontier of high-efficiency speech synthesis. In a standard integration pipeline, the traditional continuous regression heads optimized via Mean Squared Error are replaced by discrete classification networks optimized via Cross-Entropy loss (Du et al., 2024). This fundamental shift effectively eliminates spectral over-smoothing artifacts. Early integration strategies frequently treat discrete tokens as direct substitutes for continuous features, mapping linguistic hidden states directly to a single-layer vector quantization codebook to bypass complex hierarchical dependencies (Polyak et al., 2021). While computationally efficient, relying on a severely truncated codebook fundamentally limits the acoustic upper bound and degrades speech naturalness.

To accommodate the complete multi-layer residual vector quantization structure without autoregressive delays, recent methodologies have explored advanced generative paradigms. For instance, continuous latent diffusion models have been cascaded with discrete codecs to achieve expressive zero-shot generation (Ju et al., 2024; K. Shen et al., 2024). Similarly, adversarial training combined with latent style diffusion has successfully matched human-level naturalness without requiring reference audio (Y. A. Li et al., 2023). Other approaches, such as non-autoregressive flow-matching models, achieve remarkable temporal parallelism by infilling speech directly from text (Le et al., 2023), while intrinsic prosody disentanglement strategies separate global style from local acoustic tokens to improve fine-grained controllability (Jiang et al., 2023).

Although these hybrid architectures achieve exceptional acoustic fidelity, their heavy iterative sampling or multi-stage generation processes inherently contradict the primary objective of lightweight non-autoregressive models: ultra-low latency generation. Developing a fully feed-forward, single-pass integration strategy that respects the multi-layered dependencies of modern audio codecs remains an active engineering challenge.

## 3. Research Methodology

### 3.1 Overview

To support real-time human-computer interaction, this study addresses the computational bottlenecks inherent in conventional speech synthesis pipelines by proposing an end-to-end architecture optimized for block-wise generation. By directly modeling discrete latent representations, the proposed architecture achieves ultra-low latency inference without requiring complex incremental chunking mechanics, thereby satisfying strict industrial deployment constraints.

The strategic integration of a FastSpeech 2 (Ren, Hu, et al., 2022) non-autoregressive backbone with the hierarchical discrete latent space of the Mimi neural audio codec (Défossez et al., 2024) constitutes a structural engineering decision designed to satisfy strict industrial deployment constraints. Contemporary generative architectures often rely on computationally intensive flow-based networks or iterative diffusion sampling, which incur severe execution delays. Conversely, autoregressive large language models introduce prohibitive algorithmic latency that fundamentally limits continuous streaming. FastSpeech 2 provides a deterministic and fully non-autoregressive feed-forward topology that maximizes parallel temporal computation. However, standard FastSpeech 2 architectures rely on continuous high-resolution Mel-spectrograms, inherently necessitating computationally expensive neural vocoders. By coupling this lightweight backbone explicitly with the Mimi codec, the system achieves a profound structural advantage. Unlike alternative neural audio codecs operating at higher temporal resolutions, Mimi compresses acoustic information into a low frame rate of 12.5 Hertz. This temporal compression perfectly complements the quadratic computational complexity of the FastSpeech 2 transformer attention mechanisms by reducing the target sequence length by a significant factor of eight. Consequently, this precise architectural pairing mutually resolves their individual structural limitations, eliminating a critical bottleneck in streaming pipelines and enabling ultra-low latency inference for practical deployment.

The systematic operational pipeline of this real-time deployable architecture encompasses five distinct phases, as illustrated in Figure 1. During the initial data preprocessing and alignment phase, raw acoustic corpora are processed to extract linguistic phonemes via forced alignment alongside continuous auxiliary features, which are subsequently synchronized with the target 12.5 Hertz discrete codec frames. The second phase establishes the model architecture design, constructing the modified end-to-end non-autoregressive encoder-decoder architecture. This phase introduces a specialized multi-head decoder to conditionally manage the 32 hierarchical codebook sequences alongside an auxiliary projection branch designed to strictly enforce acoustic consistency. Subsequently, the training strategy phase employs a hybrid multi-task objective function to mitigate spectral over-smoothing. This robust optimization strategy combines cross-entropy loss for discrete token classification, mean squared error for continuous prosody regression, and an auxiliary L1 loss for fine-grained spectral reconstruction. During the fourth phase of inference and waveform reconstruction, the trained model processes input text to conditionally generate discrete tokens. These tokens are directly decoded into high-fidelity speech waveforms by the pre-trained Mimi neural vocoder, establishing a completely end-to-end deployment pipeline. Finally, the performance evaluation phase conducts rigorous objective metrics and subjective perceptual tests against ground truth references to comprehensively validate the acoustic fidelity, inference efficiency, and language-independent robustness of the practical deployment.

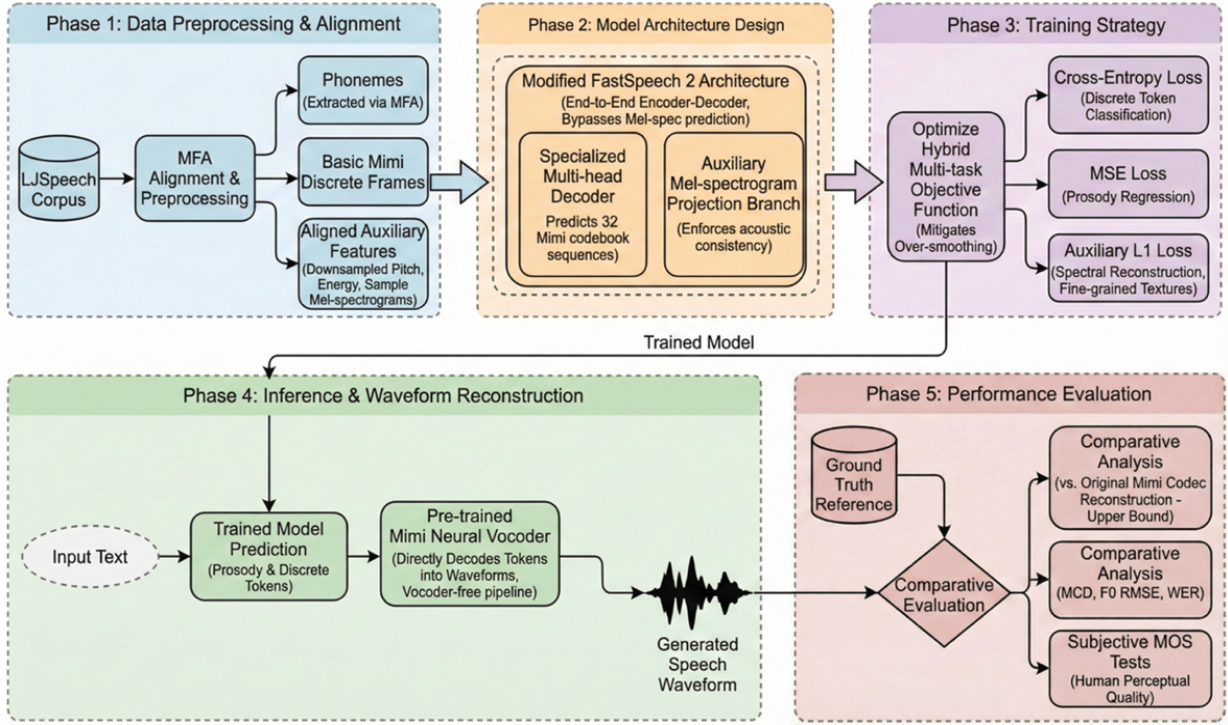

Figure 1 Research Methodology Overview

**3.2 Data Processing and Feature Alignment**

To evaluate the language-independent generalization and low-resource deployment capabilities of this resource-efficient architecture, the input text streams for both the English and Malay datasets are systematically processed into discrete phoneme sequences. For the English corpus, standard text normalization and phonemic conversion are applied. Conversely, exploiting the highly phonetic orthography of the Malay language, the Malay text undergoes a direct character-level tokenization strategy while strictly preserving explicit silence boundaries to maintain natural prosodic pauses. To resolve duration targets, both linguistic streams are processed utilizing the Montreal Forced Aligner (McAuliffe et al., 2017) equipped with language-specific pronunciation lexicons to extract precise absolute temporal boundaries. These temporal spans are subsequently multiplied by the target frame rate, yielding integer duration targets that precisely dictate the number of acoustic frames each phoneme occupies.

Establishing a rigorous temporal mapping between the discrete linguistic inputs and the continuous acoustic signals is a critical prerequisite for achieving ultra-low latency inference. All raw acoustic waveforms are uniformly resampled to 24 kHz and encoded by the pre-trained Mimi neural audio codec into multi-layered discrete target tokens utilizing residual vector quantization. This encoder operates at an 80-millisecond hop size, establishing a highly compressed 12.5 Hertz temporal resolution that eliminates a critical bottleneck in streaming pipelines. However, this extreme temporal compression introduces a severe structural challenge: transient phonemes or explicit textual boundaries frequently span less than a single 80-millisecond frame, yielding zero corresponding codec tokens. If left unaddressed, these sub-frame linguistic units fundamentally break the strict one-to-one sequence integrity required by parallel non-autoregressive length regulators, inevitably inducing severe alignment drift.

To systematically resolve this, we introduce a specialized dummy token mechanism that injects a synthetic placeholder vector for these undetected sub-frame units. By enforcing a minimum duration constraint of one frame, this structural intervention preserves the absolute sequence completeness and maintains the parallel computational integrity of the transformer backbone without introducing acoustic artifacts. During real-time execution, these dummy tokens function as seamless synchronizers that instruct the system to bypass acoustic rendering for structural boundaries, thereby guaranteeing uninterrupted generation. Simultaneously, to capture critical prosodic variations, the fundamental frequency is extracted utilizing the YIN algorithm bounded between 50 and 500 Hertz and transformed into the logarithmic domain, while frame-wise energy is computed via root-mean-square amplitude. Because these continuous variance features are initially extracted at differing sampling rates, they undergo strict linear interpolation to synchronize perfectly with the overarching 12.5 Hertz discrete codec frames, followed by a rigorous global statistical normalization excluding silent segments.

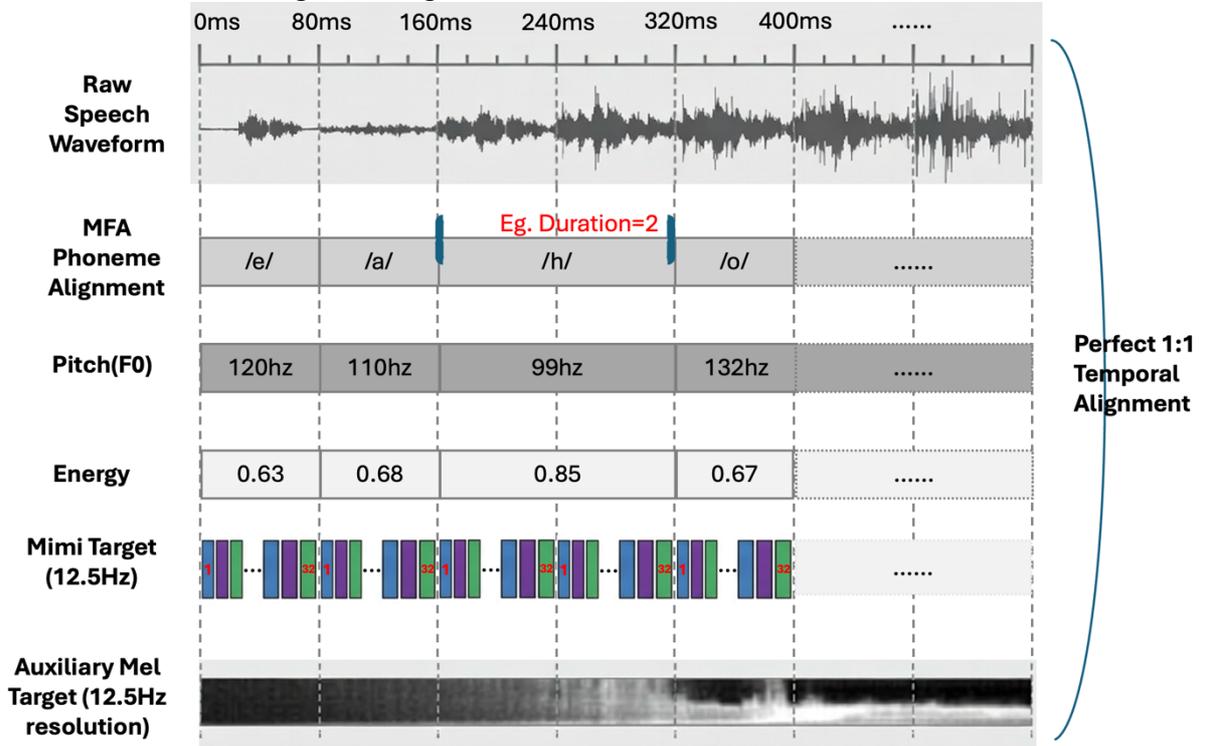

Figure 2 Data Alignment Strategy

As illustrated in Figure 2, a fundamental structural challenge during this multi-resolution alignment process is resolving the temporal mismatch between standard high-resolution continuous Mel-spectrograms extracted at an approximately 10-millisecond hop rate and the highly compressed discrete Mimi tokens operating at the 80-millisecond hop rate. To enable synchronous auxiliary spectral supervision, we implement a targeted linear interpolation downsampling strategy. Let $M_{high}$ denote the high-resolution Mel features. For the $t$-th Mimi frame, the aligned continuous acoustic feature $M_{aligned}^{(t)}$ is derived via targeted average pooling:

$$M_{aligned}^{(t)} = \frac{1}{K} \sum_{K=1}^{K} M_{high}^{(t \cdot K + k)} \qquad (1)$$

where $K = 8$ represents the precise downsampling factor required to bridge the sampling variance. Leveraging the extracted phoneme durations, the length regulator explicitly expands the corresponding linguistic hidden states alongside the aligned variance features, ensuring a strict one-to-one temporal synchronization. This deterministic alignment successfully guarantees that the auxiliary spectral regularization loss is computed synchronously with the discrete token prediction, seamlessly harmonizing continuous acoustic regression targets with discrete hierarchical codec representations. By unifying these multi-resolution features, the proposed architecture completely removes the dependency on computationally expensive neural vocoders, establishing a highly scalable foundation for practical deployment.

### 3.3 Model Architecture

Real-world interactive applications necessitate streaming architectures capable of generating audio incrementally with minimal algorithmic latency. To achieve a rigorous balance between synthesis quality and structural efficiency for practical deployment, the proposed architecture is meticulously engineered as a resource-efficient end-to-end architecture. By unifying a non-autoregressive frontend directly with a discrete neural audio codec backend, this structural design eliminates a critical bottleneck in streaming pipelines and completely removes the dependency on computationally expensive neural vocoders.

As explicitly illustrated in Figure 3, the overarching system architecture is logically partitioned into three cohesive operational modules designed to systematically facilitate ultra-low latency inference. Module 1 encapsulates the modified FastSpeech 2 backbone, serving as the primary non-autoregressive acoustic frontend to process initial discrete linguistic inputs into time-aligned continuous hidden states. Operating as the core architectural bridge, Module 2 introduces the specialized discrete token predictor to transform these continuous states into multi-layered discrete codebook tokens. Finally, Module 3 directly incorporates the pre-trained Mimi neural codec as the robust synthesis backend, exclusively processing the predicted discrete tokens to generate high-fidelity raw speech waveforms. This streamlined tripartite progression strictly guarantees real-time deployable execution while firmly preserving the structural integrity of the generated acoustic textures.

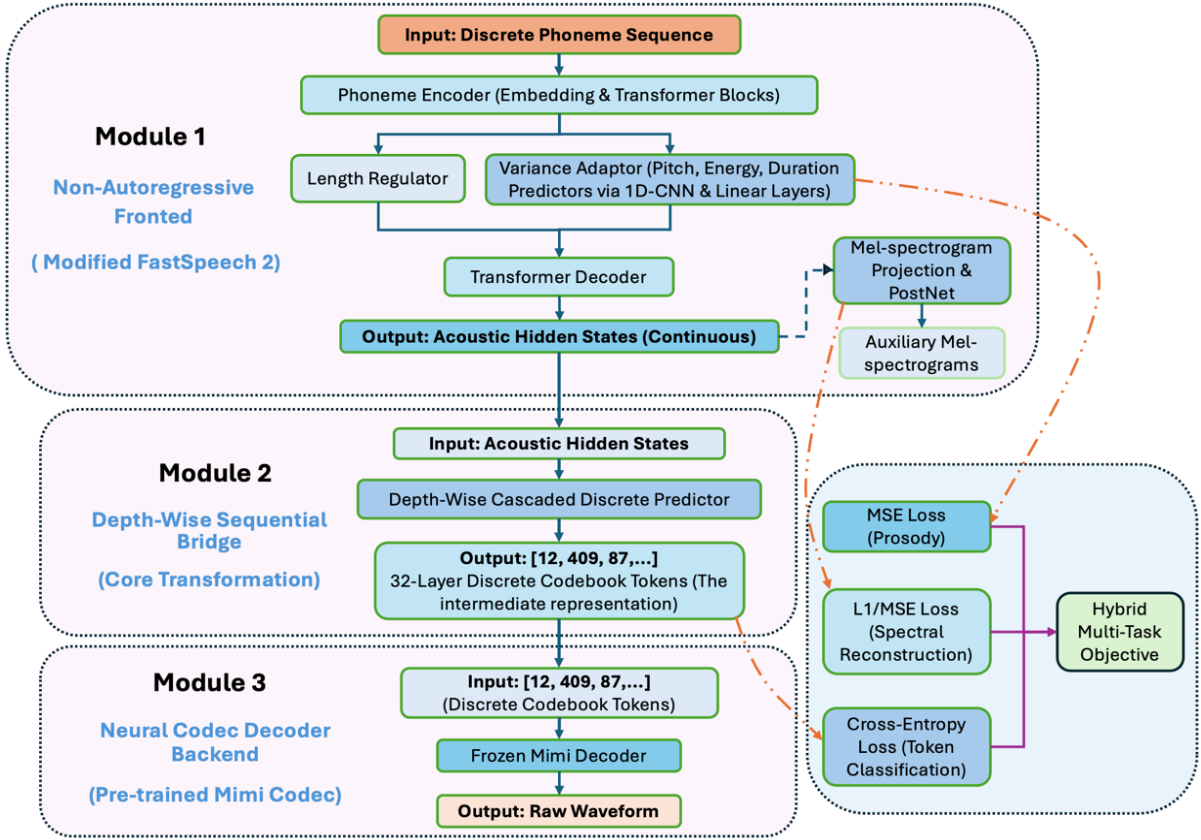

Figure 3 NAR Acoustic Codec Architecture

*3.3.1 Encoder and Variance Adaptor*

     As delineated within the first module of Figure 3, the phoneme encoder and variance adaptor constitute the fundamental components of the modified FastSpeech 2 frontend. The phoneme encoder processes the discrete linguistic tokens through an initial linear embedding layer followed by multiple feed-forward transformer blocks to construct rich semantic representations. These sequential representations are subsequently routed into the variance adaptor, which utilizes sequential one-dimensional convolutional networks and dedicated linear projection layers to accurately predict continuous prosodic features, specifically log-scale duration, fundamental frequency, and frame-wise energy. Driven by the predicted durations, the length regulator explicitly expands the linguistic hidden states to match the highly compressed 12.5 Hertz temporal resolution demanded by the discrete Mimi codec, ensuring absolute temporal synchronization prior to discrete classification.

*3.3.2 Depth-Wise Cascaded Discrete Decoder*

     To effectively model the complex hierarchical residual vector quantization structure comprising 32 distinct quantizers, the conventional single regression head is replaced by a highly specialized depth-wise cascaded discrete classifier, visually represented as the core transformation bridge in Module 2 of Figure 3. Rather than predicting all quantization layers naively in parallel, an approach that entirely disregards critical inter-layer acoustic dependencies, the proposed decoder predicts the discrete tokens sequentially across the quantization depth for each individual temporal frame.

Let $h_t$ denote the base acoustic hidden state generated by the non-autoregressive backbone at time step $t$. To preserve strictly hierarchical conditioning, the hidden state is progressively updated by accumulating the embedded representations of the predicted tokens from all preceding layers. Specifically, the conditioned hidden state $h_{t,i}$ for the $i$-th quantizer is recursively defined as:

$$h_{t,i} = h_t + \sum_{j=1}^{i-1} E_j(y_{t,i}) \tag{2}$$

where $E_j(\cdot)$ denotes the learned embedding lookup function for the $j$-th quantization codebook, and the summation represents the cumulative feature fusion. For the initial semantic layer, the conditioned state remains identical to the base state. Subsequently, the probability distribution over the codebook vocabulary containing 2048 discrete codes is computed utilizing a linear projection combined with a $softmax$ activation:

$$P(y_{t,i}|h_{t,i}) = softmax(W_i h_{t,i} + b_i) \tag{3}$$

where $W_i$ and $b_i$ represent the specific learnable weights and biases for the $i$-th quantization head.

This depth-wise sequential mechanism explicitly compels the architecture to condition higher-order fine-grained acoustic details upon the broad semantic content established by the lower quantizers. Crucially, because this progressive projection occurs strictly across the feature depth within individual frames rather than across the extended temporal sequence, it successfully restores acoustic fidelity and mitigates intelligibility degradation while absolutely preserving the ultra-low real-time factor required for high-efficiency streaming inference.

*3.3.3 Auxiliary Mel Supervision Branch*

To prevent discrete latent space collapse and strictly enforce continuous acoustic consistency during the learning phase, the architecture integrates a parallel auxiliary supervision branch, depicted alongside the primary decoder in the frontend module of Figure 3. This branch projects the expanded continuous hidden states into 80-channel Mel-spectrograms utilizing a dedicated linear layer, followed immediately by a five-layer convolutional PostNet to refine the high-frequency spectral details. Although these continuous Mel features are completely discarded during real-time inference to ensure ultra-low latency, they provide a highly stable and critical continuous regularization signal during training, effectively guiding the discrete token predictor to capture accurate spectral envelopes.

**3.4 Training Objective**

As aggregated at the bottom of the architectural diagram in Figure 3, the entire streaming architecture is optimized from end to end utilizing a hybrid multi-task objective function that seamlessly combines discrete classification and continuous regression penalties. The overarching loss function is mathematically formulated as:

$$L_{total} = L_{token} + \lambda_{dur}L_{dur} + L_{pitch} + L_{energy} + \lambda_{mel}(L_{mel} + L_{postnet}) \tag{4}$$

For the primary discrete synthesis target $L_{token}$, cross-entropy is utilized to maximize the log-likelihood of the ground-truth tokens. Aligning with the depth-wise sequential architecture, this classification loss is aggregated across all 32 hierarchical quantization steps for each temporal frame. To prevent the information collapse of higher-order codebooks, a staged weighting strategy is applied where codebooks 1 through 4 receive a full weight of 1.0, intermediate codebooks 5 through 16 are scaled by 0.5, and fine-grained codebooks 17 through 32 are weighted by 0.1.

To guarantee alignment stability and systematically prevent the structural skipping of transient phonemes, the duration penalty $L_{dur}$ measures the mean squared error between the predicted and ground-truth log-durations, formally defined as $L_{dur} = \left\| log(\hat{d}) - log(d+1) \right\|_2^2$, with the hyperparameter $\lambda_{dur}$ empirically set to 2.0. Similarly, $L_{pitch}$ and $L_{energy}$ employ standard mean squared error to optimize the prosodic variance predictors. Finally, the weighting coefficient $\lambda_{mel}$ for the continuous auxiliary spectral losses is set to 10.0, establishing an optimal gradient balance between the ultra-low latency discrete classification task and the high-fidelity continuous reconstruction regularization.

### 3.5 Inference Pipeline

During practical deployment, the proposed architecture operates as a strictly end-to-end, resource-efficient streaming architecture. The predicted phoneme durations and the injected structural dummy tokens deterministically establish the precise temporal sequence length without requiring future context buffering. For each generated acoustic frame, the cascaded discrete decoder dynamically predicts the 32 hierarchical layers of codec tokens utilizing the depth-wise sequential feature fusion mechanism. Because this progressive projection is strictly confined to the feature depth within individual acoustic frames rather than extending across the sequential time axis, it introduces negligible computational overhead and guarantees ultra-low latency inference. Finally, the resulting discrete indices are fed directly into the frozen pre-trained Mimi neural decoder to instantly reconstruct the high-fidelity 24 kHz audio waveform. This streamlined execution completely removes the dependency on computationally expensive neural vocoders and bypasses the severe algorithmic latency associated with continuous phase-estimation networks, thereby establishing a highly scalable and robust foundation for real-time deployable speech interfaces.

## 4. Experiments and Results

### 4.1 Experimental Setup

#### 4.1.1 Dataset and Preprocessing

The proposed streaming architecture is rigorously evaluated utilizing two distinct single-speaker corpora to validate its language independent structural robustness. For the English benchmark, we utilize the standard LJSpeech corpus encompassing approximately 24 hours of high-fidelity audio (Ito & Johnson, 2017). For the Malay evaluation, we employ the open-source Malaysian speech dataset provided by the Mesolitica research team (Husein & Mesolitica, 2023), specifically utilizing a 13-hour subset featuring a native female speaker.

To strictly satisfy the input resolution requirements of the Mimi neural audio codec, all raw acoustic waveforms across both datasets are uniformly resampled to 24 kHz. Dataset-specific preprocessing is subsequently applied to optimize acoustic quality. The English audio undergoes a 60-decibel threshold application for background silence trimming. Concurrently, the Malay transcripts undergo rigorous text normalization to expand numerical and currency symbols into their full word equivalents, alongside a strategic filtering of foreign loanword suffixes to maintain phonological purity and ensure the acoustic model focuses entirely on native pronunciation. Finally, to construct the parallel regularization targets, 80-channel Mel-spectrograms are computed for all utterances employing a window length of 1024 samples and a hop length of 256 samples, providing the necessary continuous spectral representations for the auxiliary supervision branch.

*4.1.2 Implementation and Training Protocol*

The end-to-end streaming architecture is implemented in PyTorch and trained on a single NVIDIA GeForce RTX 4090 GPU equipped with 24 gigabytes of memory. Network optimization is driven by the Adam optimizer utilizing beta values of 0.9 and 0.98 alongside an epsilon of $10^{-9}$ and a processing batch size of 16. The learning rate strictly follows a scheduled decay mechanism, linearly warming up over the initial 4000 steps before transitioning into a smooth exponential annealing phase.

To explicitly mitigate quantization artifacts in high-frequency bands during the depth-wise sequential decoding, the 32 hierarchical classification heads are governed by a staged loss weighting strategy. Specifically, the fundamental semantic codebooks 1 through 4 are assigned a full weight of 1.0, intermediate codebooks 5 through 16 are scaled to 0.5, and the fine-grained acoustic codebooks 17 through 32 are reduced to 0.1. Simultaneously, the auxiliary Mel-spectrogram reconstruction loss is amplified by a coefficient of 10 to firmly enforce spectral continuity. Based on this highly optimized configuration, the architecture achieves complete acoustic stability and optimal gradient convergence at 200,000 training steps for the English model. Reflecting the architectural efficiency and the compressed latent space representation, the Malay model achieves optimal synthesis quality significantly faster, reaching complete convergence within merely 90,000 training steps.

## 4.2 Objective Evaluation

*4.2.1 Acoustic Fidelity and Intelligibility*

We evaluated the proposed end-to-end streaming architecture against a theoretical topline system utilizing ground-truth audio reconstructed directly via the pre-trained Mimi codec. For comprehensive comparative baselines on the English benchmark, we selected two robust non-streaming models comprising the generative VITS and the standard FastSpeech 2 integrated with a continuous HiFi-GAN vocoder. Objective acoustic fidelity was quantified utilizing Mel-Cepstral Distortion, Band-Aperiodicity, fundamental frequency root mean square error, and Voiced/Unvoiced Error. Speech intelligibility was rigorously assessed through Word Error Rate computed by the Whisper automated speech recognition system.

As presented in Table 4, the English baseline systems achieve strong objective metrics, with VITS and FastSpeech 2 recording Mel-Cepstral Distortion scores of 7.31 decibels and 8.24 decibels respectively. The proposed method yields a competitive score of 10.20 decibels. While marginally higher than the continuous baselines, this performance reflects a highly optimized engineering trade-off when substituting high-resolution continuous Mel-spectrograms with highly compressed discrete multi-layer tokens. Most notably, the proposed architecture achieves an exceptional Voiced/Unvoiced Error of 2.67%, successfully outperforming both the FastSpeech 2 baseline at 3.62% and the VITS baseline at 2.82%. This result demonstrates that the depth-wise sequential feature fusion mechanism remarkably excels at capturing fundamental voicing characteristics despite the aggressive discrete compression.

| Model | MCD (dB) | BAP (dB) | F0 RMSE (Hz) | V/UV Error (%) |
|---|---|---|---|---|
| **Topline (Mimi Reconstruction)** | 6.27 | 6.31 | 37.85 | 2.38 |
| **VITS (Baseline)** | 7.31 | 6.80 | 48.99 | 2.82 |
| **FastSpeech 2 (Baseline)** | 8.24 | 8.14 | 57.01 | 3.62 |
| **Proposed Method (FS2 + Mimi) English** | 10.20 | 9.20 | 62.01 | 2.67 |
| **Proposed Method (FS2 + Mimi) Malay** | 11.60 | 11.45 | 41.59 | 2.06 |

Table 4 Acoustic Fidelity Comparison on LJSpeech Test Set

To further validate the language-independent structural robustness of the proposed architecture, identical objective acoustic measurements were conducted on the Malay corpus. The generated Malay speech exhibits exceptional acoustic stability, recording a fundamental frequency error of merely 41.59 Hertz and an astonishingly low Voiced/Unvoiced Error of 2.06%. These metrics strongly indicate that the discrete hierarchical modeling paradigm adapts highly efficiently to the distinct phonological structures of diverse languages, ensuring precise pitch trajectory reconstruction and voicing classification without requiring dataset-specific structural modifications.

Regarding speech intelligibility, as presented in Table 5, the English benchmark results follow a highly promising trend. The end-to-end generative VITS and the standard FastSpeech 2 baseline achieve Word Error Rates of 1.64% and 5.19% respectively. In comparison, the proposed end-to-end architecture records a Word Error Rate of 8.89%. By firmly maintaining a single-digit error rate within a highly compressed discrete space, the system successfully preserves speech intelligibility. Due to the inherent transcription variance and limited reliability of automated speech recognition systems for low-resource languages, absolute intelligibility benchmarking is restricted to the English corpus to ensure statistical rigor.

| **Model** | **WER (%)** | **Description** |
|---|---|---|
| **Topline (Mimi Reconstruction)** | 1.34 | Ground Truth→Mimi Encoder→Decoder |
| **VITS (Baseline)** | 1.64 | End-To-End Generative Model |
| **FastSpeech 2 (Baseline)** | 5.19 | Text→FastSpeech 2→HiFi-GAN Vocoder |
| **Proposed Method (FS2 + Mimi)** | 8.89 | Text→FastSpeech2→Mimi Tokens→Decoder |

Table 5 Word Error Rate (WER) Comparison

This significant single-digit achievement directly validates the architectural efficacy of the proposed depth-wise sequential decoding strategy. Unlike naive parallel prediction models that suffer from severe phonetic alignment degradation, the progressive feature fusion mechanism explicitly forces the higher-order residual quantizers to condition upon the semantic foundations established by the initial layers. This cascading temporal memory effectively prevents the structural collapse of fine-grained acoustic textures, successfully bridging the representational gap between continuous spectrogram regression and multi-layered discrete classification.

*4.2.2 Inference Efficiency (Real-Time Factor)*

A primary contribution of this work lies in the exceptional computational efficiency of the end-to-end streaming architecture across diverse linguistic inputs. We evaluated the inference speed on an NVIDIA GeForce RTX 4090 GPU utilizing randomly sampled test utterances for both the English and Malay benchmarks.

As detailed in Table 6, the end-to-end English system achieves an ultra-low Real-Time Factor of 0.0033, effectively generating audio approximately 303 times faster than real-time. Correspondingly, the Malay model records a highly efficient Real-Time Factor of 0.0055, firmly maintaining a real-time generation speedup exceeding 179-fold. The slight numerical variance in absolute latency arises naturally from the varying average sequence lengths and phoneme densities inherent to the different linguistic test sets. Nevertheless, both configurations decisively satisfy the strict ultra-low latency requirements demanded by real-time streaming applications.

| Language Benchmark | Component | Average Latency (ms) | RTF |
|---|---|---|---|
| **English** | Depth-Wise FastSpeech 2 Decoder | 8.06 | - |
| **English** | Mimi Decoder (32 codebooks) | 11.58 | - |
| **English** | **End-to-End Total** | ~19.63 | ~0.0033 |
| **Malay** | Depth-Wise FastSpeech 2 Decoder | 8.00 | - |
| **Malay** | Mimi Decoder (32 codebooks) | 13.12 | - |
| **Malay** | **End-to-End Total** | 21.12 | ~0.0055 |

Table 6 Inference Latency and Real-Time Factor across English and Malay Benchmarks

This sustained language-independent efficiency is fundamentally attributed to the complete eradication of temporal autoregressive bottlenecks. Although the proposed architecture employs a depth-wise sequential projection to meticulously preserve acoustic fidelity, this recursive operation is strictly confined to the feature depth of individual frames rather than extending across the sequential time axis. Consequently, operating entirely within the highly compressed 12.5 Hertz temporal resolution of the discrete latent space drastically reduces the quadratic computational load on the Transformer self-attention mechanisms when compared to standard 100 Hertz continuous spectrogram pipelines, thereby establishing a highly scalable paradigm for multilingual streaming deployments.

*4.2.3 Comparative Benchmarking and Computational Complexity*

To further contextualize the computational advantages of the proposed architecture, we benchmarked its absolute inference speed against three standard baseline architectures. These baselines comprise the end-to-end generative VITS model alongside a conventional continuous FastSpeech 2 acoustic model coupled with heavy neural vocoders including HiFi-GAN and Parallel WaveGAN.

| System | Acoustic Model | Vocoder | Overall RTF |
|---|---|---|---|
| **Baseline 1** | FastSpeech 2 | HiFi-GAN | ~0.025 |
| **Baseline 2** | FastSpeech 2 | Parallel WaveGAN | ~0.035 |
| **End-to-End** | VITS | None (End-to-End) | ~0.020 |
| **Model Method** | FastSpeech 2 (Modified) | Mimi | ~0.0033 |

Table 7 Comparative Inference Speed across TTS Systems

As illustrated in Table 7, the proposed end-to-end architecture demonstrates a monumental leap in system efficiency, achieving a 6-fold speedup over the highly optimized VITS architecture and up to a 10.6-fold speedup compared to the cascaded continuous pipelines. Standard continuous vocoders and flow-based generative models inherently rely on deep multi-receptive field convolutional networks utilizing dense transposed convolutions or complex adversarial operations to upsample features into high-resolution audio. These operations consume massive floating-point computations and inevitably create a severe execution bottleneck.

In contrast, the proposed method directly maps linguistic features into a compact discrete latent space. By leveraging the highly optimized pre-trained Mimi decoder to reconstruct waveforms directly from the 32 layers of residual vector quantization tokens, the system entirely circumvents the computationally prohibitive continuous phase-estimation process. This comparative advantage confirms that the proposed depth-wise architecture is exceptionally well-suited for deployment in resource-constrained streaming environments where minimizing algorithmic latency is paramount.

*4.2.4 Time-to-First-Byte Latency Analysis*

Algorithmic latency is evaluated utilizing the Time-to-First-Byte metric on a single NVIDIA RTX 4090 GPU. In the context of the established chunk-level streaming paradigm, this metric represents the absolute wall-clock time measured from the input of a pre-segmented linguistic chunk to the synthesis of the initial playable audio waveform. The proposed discrete architecture achieves an average Time-to-First-Byte of 48.99 milliseconds alongside a 90th percentile latency of 22.03 milliseconds across the test evaluations. Because conventional continuous baselines fundamentally necessitate generating the entire utterance spectrogram before initiating the computationally dense vocoder phase, establishing a direct millisecond-level comparative baseline under identical streaming constraints is structurally impractical. Consequently, the critical evaluation criterion for streaming viability is the human perception threshold. The recorded average delay of 48.99 milliseconds falls significantly below the widely accepted 200-millisecond threshold required for imperceptible latency in real-time conversational interfaces. This empirical measurement confirms that the depth-wise sequential decoding mechanism resolves discrete acoustic artifacts while successfully maintaining the rigorous speed requirements demanded by interactive streaming deployments.

**4.3 Subjective Evaluation**

To comprehensively assess perceptual naturalness and acoustic fidelity, we conducted a blind Mean Opinion Score evaluation involving 11 proficient listeners. The evaluation encompassed both the English and Malay test sets, where listeners rated the synthesized utterances against human ground truth recordings on a standard five-point scale. To ensure strict statistical rigor, all results are reported alongside their 95% confidence intervals.

As detailed in Table 8, the English evaluation compares the proposed streaming architecture against the continuous FastSpeech 2 pipeline coupled with a Parallel WaveGAN vocoder. The proposed method yields a score of 2.51, compared to 2.81 for the baseline. While a perceptual gap exists, this outcome reflects a deliberate and highly optimized efficiency-quality trade-off. The baseline achieves its score entirely by relying on high-resolution continuous Mel-spectrograms and a computationally dense neural vocoder. Conversely, the proposed discrete architecture completely eliminates the vocoder bottleneck to enable ultra-low latency streaming, generating intelligible and functionally viable speech under strict computational constraints.

Crucially, the subsequent evaluation on the Malay dataset demonstrates exceptional language-independent generalization. The proposed architecture achieves an outstanding perceptual score of 4.31, closely approaching the human ground truth score of 4.65. This remarkable performance jump indicates that the discrete hierarchical modeling paradigm is exceptionally well-suited for the highly phonetic orthography and distinct phonological structure of the Malay language.

| Language benchmark | System | MOS ($\pm$ 95% CI) | Description |
|---|---|---|---|
| **English** | Ground Truth (LJSpeech) | 4.51 $\pm$ 0.08 | Original high-fidelity recordings |
| **English** | Baseline (FS2 + PWG) | 2.81 $\pm$ 0.13 | Non-streaming, Continuous Mel-spec + Vocoder |
| **English** | Proposed Method (FS2 + Mimi) | 2.51 $\pm$ 0.11 | Streaming, Discrete Tokens, End-to-End |
| **Malay** | Ground Truth (Melsolitica) | 4.65 $\pm$ 0.08 | Original high-fidelity recordings |
| **Malay** | Proposed Method (FS2 + Mimi) | 4.31 $\pm$ 0.11 | Streaming, Discrete Tokens, End-to-End |

Table 8 Subjective Evaluation (MOS) Results with 95% Confidence Intervals

A discrepancy is observed between the subjective perceptual quality of the English benchmark and the Malay benchmark. This efficiency-quality divergence stems from three primary interacting factors. First, English features a deep orthography with highly complex pronunciation rules requiring extensive global semantic context to maintain prosodic naturalness, whereas the Malay language utilizes a transparent phonetic orthography. This direct character-to-phoneme mapping complements localized chunk-level discrete prediction, enabling the non-autoregressive architecture to perform efficiently on the Malay dataset. Second, subjective evaluation bias significantly influences the scoring distribution. The English mean opinion scores were evaluated by non-native volunteer listeners who frequently demonstrate divergent standards regarding minor phonetic artifacts and machine-like accents. Conversely, the Malay evaluations were conducted by native speakers, providing a more calibrated assessment of natural speech distributions. Finally, the inherent acoustic stylistic variations between the datasets contribute to the performance gap. The English LJSpeech corpus comprises audiobook readings characterized by dramatic prosodic variance, which presents a substantially higher modeling difficulty for aggressively compressed 12.5 Hertz discrete architectures compared to the relatively consistent acoustic conditions of the Malaysian conversational dataset. By reconstructing native prosody and fine-grained acoustic textures without requiring a continuous phase-estimation network, the system validates its capability for deploying multilingual streaming interfaces.

**4.4 Ablation Study**

To rigorously evaluate the structural contributions of the proposed architecture and explicitly isolate specific algorithmic variables, we conducted comprehensive ablation studies focusing on codebook depth selection, decoding strategies, and text processing methodologies. To ensure strict experimental control and minimize redundant computational overhead, all subsequent ablation evaluations were executed exclusively on the English benchmark.

*4.4.1 Impact of Quantization Layer Depth*

We investigated the structural impact of the residual vector quantization layer depth on the overall synthesis quality under the adopted depth-wise sequential decoding architecture. As detailed in Table 9, we compared the acoustic performance of a truncated representation utilizing 16 codebooks against the full hierarchical set comprising 32 codebooks.

| Decoding Strategy | WER (%) | MCD (dB) | BAP (dB) | F0 RMSE (Hz) | V/UV Error (%) |
|---|---|---|---|---|---|
| **16 Codebooks (Truncated)** | 9.07 | 10.38 | 9.28 | 61.81 | 2.84 |
| **32 Codebooks (Proposed)** | 8.89 | 10.20 | 9.20 | 62.01 | 2.67 |

Table 9 Performance Comparison between 16 and 32 Codebooks decoding strategies

The objective metrics reveal that truncating the quantization depth explicitly degrades both speech intelligibility and overall spectral fidelity. The full 32-layer architecture achieves a lower Word Error Rate of 8.89% and a superior Mel-Cepstral Distortion of 10.20 dB compared to the truncated 16-layer configuration. Although the simplified 16-layer model exhibits a marginally lower fundamental frequency error due to reduced generative variance in the truncated latent space, it fundamentally fails to reconstruct critical high-frequency acoustic details. The empirical results demonstrate that the deeper residual layers are structurally crucial for resolving fine-grained acoustic textures and disambiguating complex phonetic transitions. Consequently, the complete 32-layer configuration is established as the optimal architectural baseline, successfully maximizing acoustic fidelity without violating the strict computational latency budget of the streaming system.

*4.4.2 Depth-Wise Sequential versus Naive Parallel Decoding*

To validate the critical necessity of the proposed progressive feature fusion mechanism, we conducted an ablation study comparing the depth-wise sequential decoding strategy against a naive fully parallel projection. In the naive parallel configuration, the multi-head decoder is forced to predict all 32 hierarchical codec tokens simultaneously directly from the base acoustic hidden state without any inter-layer conditioning.

As presented in Table 10, this naive parallel approach suffers from severe performance degradation, recording a Word Error Rate of 14.37% alongside a Mel-Cepstral Distortion of 12.49 dB. This degradation fundamentally occurs because simultaneous prediction entirely discards the vital conditional acoustic dependencies existing between the coarse semantic codebooks and the fine-grained detail codebooks, inevitably leading to phonetic alignment collapse and compromised acoustic fidelity.

| Decoding Strategy | WER (%) | MCD (dB) | BAP (dB) | F0 RMSE (Hz) | V/UV Error (%) |
|---|---|---|---|---|---|
| **Naive Parallel** | 14.37 | 12.49 | 12.72 | 68.69 | 3.25 |
| **Depth-Wise Sequential (Proposed)** | 8.89 | 10.20 | 9.20 | 62.01 | 2.67 |

Table 10 Performance Comparison between Decoding Strategies (32 Codebooks)

Conversely, the adopted depth-wise sequential strategy resolves this structural bottleneck by dynamically accumulating the embedded representations of preceding layers. This recursive conditioning successfully restores the critical inter-layer dependencies, drastically reducing the Word Error Rate to 8.89% and substantially improving all objective acoustic metrics. Most importantly, because this sequential integration is strictly constrained within the feature depth of individual frames rather than across the extended temporal sequence, the system achieves these significant acoustic improvements while safely preserving the ultra-low latency requirements of the streaming architecture.

*4.4.3 Impact of Subword Aggregation on Discrete Acoustic Modeling*

To address the occasional omission of extremely short phonemes by the non-autoregressive duration predictor, we investigated a subword aggregation strategy. We augmented the input vocabulary with 20 high-frequency merged phonetic units derived via a standard byte-pair encoding algorithm. The primary hypothesis was that forcing the model to allocate an expanded temporal window for an aggregated unit would mitigate the skipping of transient acoustic events.

However, empirical evaluations revealed that this structural modification precipitates a catastrophic acoustic collapse. The synthesized audio exhibited severe high-frequency mechanical artifacts and a complete loss of speech intelligibility, fundamentally precluding the extraction of meaningful objective metrics.

This catastrophic degradation originates from a fundamental architectural conflict between semantic aggregation and acoustic physical continuity. In the proposed architecture, the length regulator expands distinct hidden states to match precise phoneme durations. Aggregating disparate phonetic elements forces the network to generate a singular averaged hidden representation spanning the entire combined duration. For instance, merging a high-energy plosive with a zero-energy pause obliterates their distinct physical boundaries, yielding a structurally ambiguous and over-smoothed continuous feature.

Crucially, this ambiguous continuous representation severely disrupts the proposed depth-wise sequential decoding mechanism. An inaccurate discrete token prediction at the primary semantic quantizer subsequently operates as a flawed conditional input for the remaining residual layers. Because the network architecture recursively accumulates these embedded representations, the initial ambiguity triggers an exponential cascading error across the 32-layer deep codec network. This structural prediction collapse physically manifests as the observed periodic mechanical noise.

Consequently, this ablation study demonstrates that maintaining strict phoneme-level granularity is a mandatory architectural prerequisite. While subword aggregation accelerates semantic processing in large language models, applying it to end-to-end streaming architectures destroys physical singularity and severely compromises the mapping of continuous linguistic features to hierarchical discrete audio codecs.

### 4.5 Qualitative Spectrogram Analysis

To visually validate the acoustic fidelity improvements across languages and corroborate the objective metrics, we conducted a qualitative analysis of the generated Mel-spectrograms for both English and Malay. Figure 4 provides a compelling visual demonstration of the over-smoothing mitigation achieved by the proposed architecture. While the synthesized utterances exhibit slight temporal duration variations reflecting the dynamic nature of the independent non-autoregressive duration predictors, their spectral textures differ profoundly.

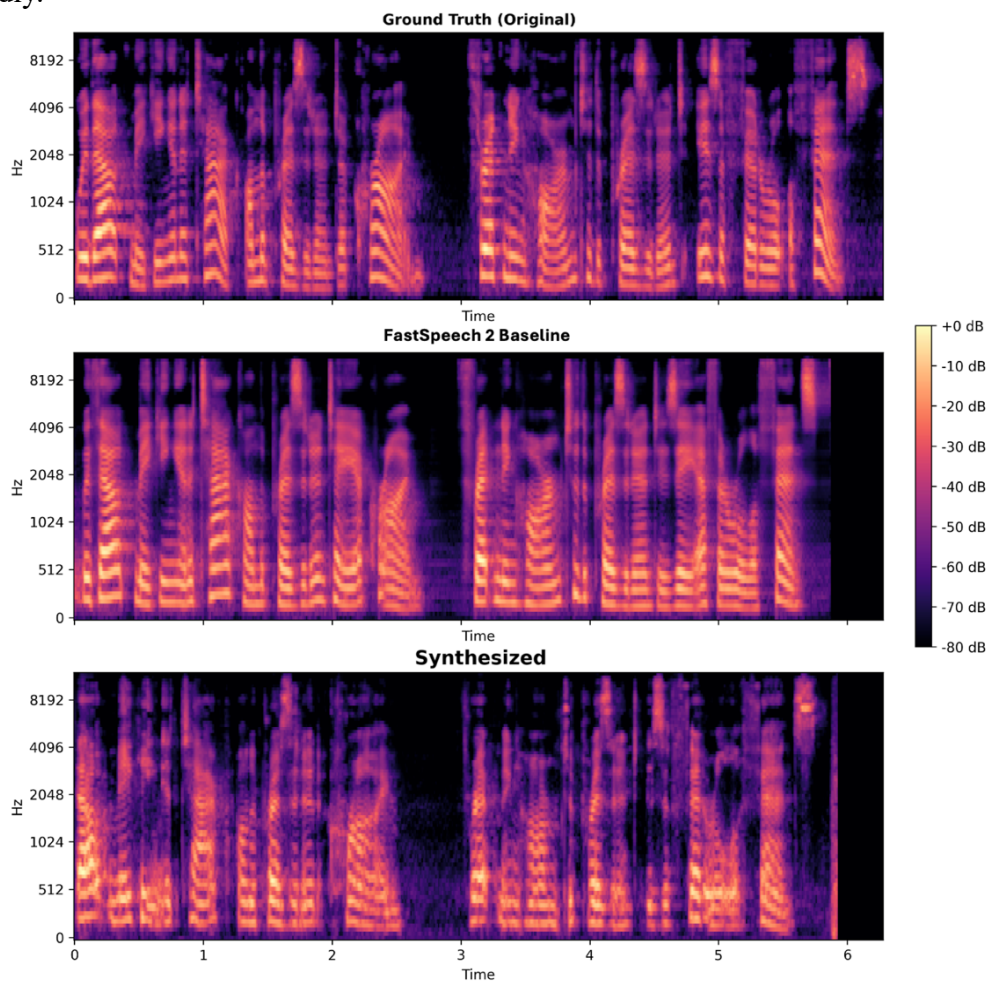

Figure 4 Qualitative Comparison of Mel-Spectrograms -- English

In the high-frequency bands above 4096 Hz, the traditional continuous FastSpeech 2 baseline exhibits severe spectral over-smoothing, rendering unvoiced fricatives and transient phonetic details as a blurred and degraded continuum. Conversely, the proposed depth-wise sequential architecture successfully recovers these fine-grained high-frequency textures and preserves clear formant transitions. This visual evidence perfectly demonstrates that the progressive discrete hierarchical modeling effectively prevents spectral degradation without requiring computationally heavy phase-reconstruction networks, thereby ensuring high-fidelity acoustic output within a streaming context.

Crucially, as illustrated in Figure 5, this superior high-frequency reconstruction capability is successfully translated to the Malay language evaluation. Similar to the English benchmark, the proposed architecture brilliantly restores the vertical striations and granular textures observed in the Malay ground truth, demonstrating exceptionally robust language-independent acoustic modeling and firmly establishing its suitability for high-quality, real-time streaming deployments in diverse resource environments.

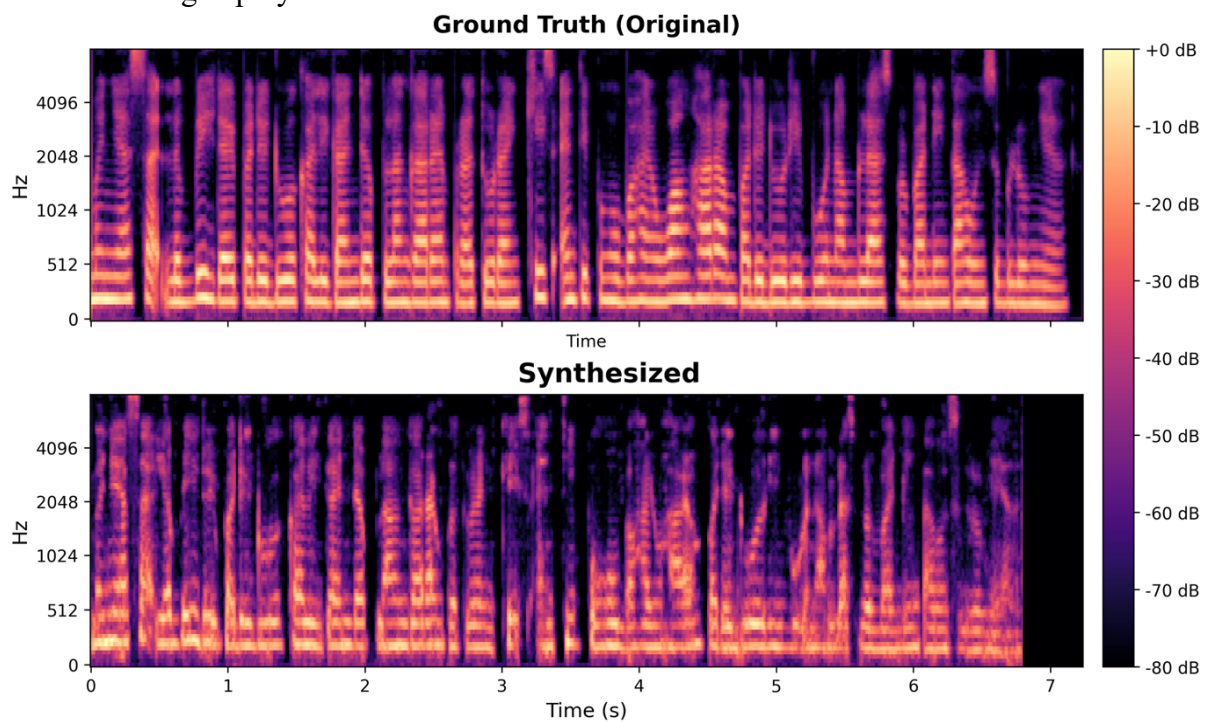

Figure 5 Qualitative Comparison of Mel-Spectrograms -- Malay

## 5. Conclusion and Future Work

### 5.1 Conclusion

This study introduces a highly efficient and end-to-end streaming architecture explicitly designed to overcome the computational bottlenecks and spectral degradation inherent in continuous acoustic modeling. By mapping linguistic features directly into a highly compressed discrete hierarchical latent space, the architecture successfully completely eliminates the dependency on computationally intensive phase-reconstruction networks. Crucially, we proposed a novel depth-wise sequential decoding strategy that dynamically conditions the multi-layered residual vector quantization codes. This progressive feature fusion mechanism effectively prevents the phonetic alignment collapse typically observed in naive parallel prediction models, firmly preserving fine-grained acoustic textures and fundamental voicing characteristics without introducing temporal autoregressive delays.

Extensive empirical evaluations conducted across English and Malay benchmarks conclusively demonstrate the immense language-independent structural robustness of the proposed paradigm. The system achieves a remarkable real-time factor speedup exceeding 179-fold across diverse phonological environments while maintaining highly competitive speech intelligibility and superior high-frequency spectral fidelity compared to traditional continuous baselines. This optimized architectural trade-off establishes a formidable and scalable foundation for deploying ultra-low latency speech generation systems in resource-constrained interactive applications.

## 5.2 Future Work

While this research establishes a robust operational foundation for streaming discrete speech synthesis, several avenues for architectural expansion remain highly promising. Despite the exceptional inference efficiency, the aggressive quantization process inherent in discrete audio codecs introduces minor structural constraints, occasionally yielding marginal quantization artifacts during highly expressive prosodic variations. Future research will focus on integrating extremely lightweight flow-matching mechanisms directly within the discrete latent space to further refine acoustic naturalness without violating the strict real-time generation budget. Additionally, exploiting the established language-independent stability of the current architecture, subsequent investigations will aim to extend this end-to-end streaming architecture to support highly scalable multi-speaker and zero-shot voice cloning capabilities across diverse linguistic domains.

**Acknowledgements:** This project has received funding from the European Union's Horizon 2020 research and innovation program under the Marie Skłodowska-Curie grant agreement No. 101007666.

**Data Availability Statement:** The datasets analyzed during the current study are publicly available. The English benchmark utilizes the LJSpeech repository, accessible at https://keithito.com/LJ-Speech-Dataset/. The Malay benchmark utilizes the single-speaker "Idayu" subset from the "Malaysian-TTS-v2" dataset curated by Mesolitica, available via the official Hugging Face repository at https://huggingface.co/datasets/mesolitica/Malaysian-TTS-v2. Custom code and models are not publicly available at this time but may be obtained from the corresponding author upon reasonable request.